\newcounter{myctr}
\def\myitem{\refstepcounter{myctr}\bibfont\noindent\ifnum\themyctr>9\else\phantom{0}\fi\hangindent17pt\themyctr.\enskip}

\documentclass{ws-ijqi}

\usepackage{amsmath}

\usepackage{amsfonts,amssymb}

\begin{document}

\markboth{Gennady P. Berman and Alexander I. Nesterov}
{Non-Hermitian adiabatic quantum optimization }

\catchline{}{}{}{}{}

\title{NON-HERMITIAN ADIABATIC QUANTUM OPTIMIZATION  }

\author{GENNADY P.  BERMAN}

\address{Theoretical Division, MS B213, Los Alamos National Laboratory\\
Los Alamos, New Mexico 87544, USA\\
gpb@lanl.gov}

\author{ALEXANDER I. NESTEROV}

\address{Departamento de F{\'\i}sica, CUCEI, Universidad de Guadalajara,
Av. Revoluci\'on 1500\\
 Guadalajara,  Jalisco,  44420, M\'exico\\
nesterov@cencar.udg.mx}

\maketitle

\begin{history}
\received{Day Month Year}
\revised{Day Month Year}
\end{history}

\begin{abstract}
We propose a novel non-Hermitian adiabatic quantum optimization algorithm. One of the new ideas is to use a non-Hermitian auxiliary ``initial''  Hamiltonian that provides an effective level repulsion for the main Hamiltonian. This effect enables us to develop an adiabatic theory which determines ground state much more efficiently than Hermitian methods.
\end{abstract}
 \keywords{critical points; ground state; quantum theory;  adiabatic quantum computation; quantum annealing}



\section{Introduction}	

Many physical and combinatorial problems associated with complex networks
of interacting degrees of freedom can be mapped to equivalent problems of
finding the ground (or minimum cost) state of a corresponding quantum
Hamiltonian $\mathcal H_0$ \cite{FGGLL,KN,SSMO,DC,SMTC,SST,CFS,SNS,AM}.
One of the approaches to finding the ground state of  $\mathcal H_0$ is adiabatic quantum computation which can be formulated as follows. Consider the time dependent Hamiltonian
$\mathcal H(t) = ({t}/{\tau})\mathcal H_0 +   (1- {t}/{\tau})\mathcal H_1$,
where $\mathcal H_0$ is the Hamiltonian whose ground state is to be found, $\tau$ is the given time-interval of quantum computation, $\mathcal H_1$ is an auxiliary ``initial" Hamiltonian and $[\mathcal H_0,\mathcal H_1] \neq 0$. As time varies from $t = 0$ to $t = \tau$, the Hamiltonian interpolates between  $\mathcal H_1$ and  $\mathcal H_0$.

If the system is initially close to the ground state of $\mathcal H_1$, and if $\tau$ is sufficiently large (slow variation), then the system will remain close to the instantaneous ground state (i.e., that of $\mathcal H_\tau(t)$) for all $t \in [0,\tau ]$. In particular, at $t = \tau$ the ground state of the total Hamiltonian, $\mathcal H_\tau$ , will be close to the ground state of $\mathcal H_0$, which is the state we seek. In practice, $\mathcal H_1$ is chosen such that its ground state is known, then the dynamics is allowed to evolve and the state of the system evolves into the final state which is the solution to the problem.

Closely related to the adiabatic quantum computation is quantum annealing, in which one has a Hamiltonian $\mathcal H_0$ to be optimized, and an auxiliary (non-commuting) term $\mathcal H_1$ so that the total Hamiltonian reads $\mathcal H_{tot}(t) = \mathcal H_0 +  \Gamma(t)\mathcal H_1$, where $\Gamma(t)$ reduces from very high value to zero during the evolution. The coefficient $\Gamma(t)$ is the control parameter and initially $\Gamma$ kept very high so that $\mathcal H_1$ dominates over $\mathcal H_0$. One starts with the ground state of $\mathcal H_1$ as the initial state, and if $\Gamma(t)$ is slowly decreasing, the adiabatic theorem guarantees reaching of the ground state of the main Hamiltonian $\mathcal H_0$ at the end of computation, assuming that there are no energy level crossings between the ground and excited states.

Quantum optimization algorithms usually require the presence of a gap between the ground state and first excited state, however, in typical cases the minimal gap, $g_m$, is exponentially small \footnote{For instance, in the commonly used quantum optimization $n$-qubit models, the estimate of the minimal energy gap is $g_m \propto 2^{-n/2}$ \cite{FGGLL,DC,SUD,JKKM,YKS}.}. This increases drastically the total computational time and from a practical point of view, the advantage of the method is lost.

In this paper, we propose a novel adiabatic non-Hermitian quantum optimization algorithm.  The non-Hermitian Hamiltonians naturally appear when a quantum system has both discrete (intrinsic) and continuous spectra, and one performs a projection of the total wave function on the discrete part of the spectrum \cite{FU,RI,RI2,VZ,KE,CIZB,KND}. In this case, the corresponding intrinsic energy levels acquire the finite widths, which are associated with the transitions from the intrinsic states to the continuum. Then the dynamics of the intrinsic states can be described by the Schr\"odinger equation with an effective non-Hermitian Hamiltonian \cite{KE,KND,CWZL,CT,DM}. The adequate candidates for implementation of non-Hermitian adiabatic quantum optimization algorithm can be, for example, superconducting phase qubits \cite{SAM,SMC,YHC,KNAB,MJM,BN1}.

We show that coupling the system to a non-Hermitian auxiliary initial Hamiltonian induces an effective level repulsion for the total Hamiltonian. This effect enables us to develop an adiabatic theory without the usual gap restriction, and to determine much more efficiently the ground state of $\mathcal H_0$.

\section{Adiabatic Quantum Optimization}

The generic adiabatic quantum optimization problem may be formulated as follows \cite{SSMO}. Let $\mathcal H_0$ be a Hamiltonian whose ground state is to be found, and $\mathcal H_1$ be the auxiliary ``initial''  Hamiltonian. Then we consider the following time -dependent Hamiltonian:
\begin{equation}\label{QA1}
\mathcal H_\tau(t) = f_0(t)\mathcal H_0 +  f_1(t)\mathcal H_1,
\end{equation}
where $[\mathcal H_0,\mathcal H_1] \neq 0$ . The functions $f_0(t)$ and $f_1(t)$ are monotonic increasing and decreasing, respectively, and satisfy the following conditions: $f_1(\tau) = 0$, $f_0(\tau) = 1$, and $f_1(0) \gg f_0(0)$, if $f_0(0)\neq 0$. The Hamiltonian $\mathcal H_\tau(t) \to \mathcal H_0$, as $t\to \tau$, and we assume that, for any choice of the function $f_0(t)$, $\mathcal H_\tau(t)$ is dominated by $\mathcal H_1$ at the initial time $t=0$ .

The evolution of the system is determined by the Schr\"odinger equation
\begin{align}\label{Sch1}
i\frac{\partial }{\partial t}|\psi(t)\rangle = { \mathcal H_\tau}(t)|\psi(t)\rangle.
\end{align}
We impose the initial conditions as follows: ${ \mathcal H_1}|\psi_g\rangle= E_g|\psi_g\rangle$, where $E_g$ denotes the energy of the ground state $|\psi_g\rangle=|\psi(0)\rangle$, which is assumed to be the ground state of the auxiliary Hamiltonian $\mathcal H_1$. The adiabatic theorem guarantees that the initial state $|\psi_g\rangle$ evolves into the final state $|\psi_g(\tau)\rangle$, which is the ground state of the Hamiltonian $\mathcal H_0$ as long as the instantaneous ground state of $\mathcal H_\tau(t)$ does not become degenerate at any time.

The validity of the adiabatic theorem requires that the following condition be satisfied \cite{SSMO,DC}:
\begin{equation}\label{QA2}
    \sum_{m\neq n}\bigg|\frac{\langle \psi_m (t)|\partial  \mathcal H_\tau/\partial
    t|\psi_n(t)\rangle}{(E_m(t) - E_n(t))^2}\bigg|\ll 1,
\end{equation}
where $|\psi_n(t)\rangle$ is the ground  instantaneous state related to the instantaneous  eigenenegry $E_n(t)$.

The condition (\ref{QA2}) is violated near the degeneracies in which the eigenvalues coalesce. In most common case of double degeneracy with two linearly independent eigenvectors, the energy surfaces form the sheets of a double cone. The apex of the cones is called a ``diabolic point'', and since, for a generic Hermitian Hamiltonian, the co-dimension of the diabolic point is three, it can be characterized by three parameters  \cite{B0,BW}. Moreover, in the vicinity of the degeneracy point, the $N$-dimensional problem effectively becomes a two-dimensional problem \cite{Arn,KMS,SKM}. This will be essential in the following argument.

For quantum optimization the commonly used version of the adiabatic theorem takes the form \cite{SSMO,DC,RN}
\begin{equation}\label{QA3}
    \tau \gg  \frac{\max |\langle \psi_e (t)|\mathcal {\dot H_\tau}(t)|\psi_g(t)\rangle|}{\min|E_e(t) - E_g(t)|^2},
\end{equation}
where $|\psi_g(t)\rangle$ and $|\psi_e(t)\rangle$ are the ground  instantaneous state and the first excited state of the total system, and ``dot'' denotes the derivative with respect to the dimensionless time $s=t/\tau$, $\mathcal {\dot H_\tau} = d\mathcal {H_\tau}/ds = \tau d\mathcal {H_\tau}(t)/dt$.

As can be observed, if during the evolution the gap $\Delta E(t) = |E_e(t) - E_g(t)|$ becomes small enough, the amount of time required to pass from the initial state to the final state becomes very large and from the practical point of view the adiabatic quantum optimization loses its advantage compared with thermal annealing.

Since in the vicinity of the level crossing point only the two-dimensional Jordan block related to the level crossing makes the most considerable contribution to the quantum evolution, the $N$-dimensional problem can be described by an effective two-dimensional Hamiltonian which can be obtained as follows \cite{NAI}. Let $t_c$ be the crossover point at which the energy gap between the ground state and the first excited state of the total Hamiltonian $\mathcal H_{\tau}(t)$ achieves its minimum. In the two-dimensional subspace corresponding to $E_g(t_c)$ and $E_e(t_c)$, we choose an orthonormal basis  $\{|0\rangle,|1\rangle\}$ and complement it to the complete basis of the $N$-dimensional Hilbert space by adding the eigenvectors  $|\psi_k (t_c) \rangle$ $(k=2,\dots,N-1)$.

Now, an arbitrary state $|\psi(t)\rangle$ can be expanded as
\begin{equation}\label{Eq1}
|\psi(t)\rangle = \alpha(t) |0\rangle + \beta (t)|1\rangle + \sum^{N-1}_{k\neq 0,1}c_k(t)|\psi_k (t_c) \rangle.
\end{equation}
Inserting this expansion into the Schr\"odinger equation (\ref{Sch1}), we obtain the coefficients, $\alpha(t)$ and $\beta(t)$, as the solution of the two-dimensional Schr\"odinger equation
\begin{align}\label{Sch2}
i \frac{\partial }{\partial t}|u(t)\rangle = \mathcal H_{ef}(t)|u(t)\rangle,
\end{align}
where
\begin{equation}\label{H1}
 \mathcal H_{ef}(t) =\left(
  \begin{array}{cc}
    \lambda  (t) + Z(t) & X(t)-iY(t) \\
   X(t)+iY(t) & \lambda(t) - Z(t),
  \end{array}
\right)
\end{equation}
and $|u(t)\rangle = \bigg (
  \begin{array}{c}
    \alpha(t) \\
    \beta (t)\\
  \end{array}\bigg)$. The matrix elements in Eq. (\ref{H1}) are determined by
\begin{align}
\lambda(t)= \frac{1}{2}(\langle 0|\mathcal H_{\tau}(t)|0\rangle + \langle 1|\mathcal H_{\tau}(t)|1\rangle) \label{R1a},\\
X(t)= \frac{1}{2}(\langle 0|\mathcal H_{\tau}(t)|1\rangle + \langle 1|\mathcal H_{\tau}(t)|0\rangle), \\
Y(t)= \frac{i}{2}(\langle 0|\mathcal H_{\tau}(t)|1\rangle - \langle 1|\mathcal H_{\tau}(t)|0\rangle ),\\
Z(t)= \frac{1}{2}(\langle 0|\mathcal H_{\tau}(t)|0\rangle- \langle 1|\mathcal H_{\tau}(t)|1 \rangle )\label{R1b}.
\end{align}

Solving the characteristic equation for $\mathcal H_{ef}(t)$, we obtain (below, we do not indicate in some expressions the explicit dependencies on t)
\begin{equation}
E_{\pm}= \lambda \pm \sqrt{X^2 + Y^2 + Z^2}.
\end{equation}
Setting $\boldsymbol R=(X,Y,Z)$, we find the energy gap between the ground state and the first excited state is $\Delta E= 2R$.

Inserting $\mathcal H_\tau(t) = f_0(t)\mathcal H_0 +  f_1(t)\mathcal H_1$ into Eqs. (\ref{R1a})  -- (\ref{R1b}), we can write the effective Hamiltonian as
\begin{equation}\label{EH1}
  \mathcal H_{ef}(t)= \lambda (t){1\hspace{-.125cm}1} +f_0(t)\boldsymbol R_0\cdot \boldsymbol \sigma + f_1(t)\boldsymbol R_1\cdot \boldsymbol \sigma,
\end{equation}
where $ \lambda (t)= f_0(t)\lambda_0 + f_1(t)\lambda_1 $, $\boldsymbol R_0 = (X_0,Y_0, Z_0)$ and $\boldsymbol R_1 = (X_1,Y_1, Z_1)$. The time-independent parameters $\lambda_0,\lambda_1$ and components of the vectors  $\boldsymbol R_0$ and $\boldsymbol R_1$ are determined from Eqs. (\ref{R1a})  -- (\ref{R1b}) by substitution of $\mathcal H_0$ and $\mathcal H_1$ instead of $\mathcal H_\tau$. Next, setting $J=f_0(t)R_0$ and $g=f_1(t)R_1$, we obtain
$R =  \sqrt{g^2 - 2g J\cos\alpha +J^2}$,
where $\cos\alpha =- \boldsymbol R_0 \cdot\boldsymbol R_1/R_0R_1$.

At the crossover point we obtain
\begin{eqnarray}\label{DP1}
g_c= -J_c\frac{{\dot J}_c - {\dot g}_c \cos\alpha}{{\dot g}_c - {\dot J}_c \cos\alpha},
\end{eqnarray}
where we denote $g_c = g(t_c)$, $J_c=J(t_c)$, $\dot g_c = \dot g(t_c)$ and $\dot J_c=\dot J(t_c)$. This yields
\begin{eqnarray}\label{DP2}
|\Delta E|_{\min}= \frac{\sqrt{{\dot g}^2_c -2 {\dot g}_c {\dot J}_c\cos\alpha + {\dot J}^2_c}}{|{\dot g}_c - {\dot J}_c \cos\alpha|}2|J_c| \sin\alpha .
\end{eqnarray}
It follows that $\sin\alpha \approx g_m /2|J_c|$, where $g_m = |\Delta E|_{\min}$  is  the minimum gap between the first two energy levels of the total Hamiltonian $\mathcal H_\tau(t)$.\\

\section{\em Non-Hermitian Adiabatic Quantum Optimization}

Non-Hermitian quantum optimization can be implemented by the following generalization of the Hermitian adiabatic quantum optimization:
\begin{equation}\label{NQA4}
 {\mathcal {\tilde H}_\tau(t)} = f_0(t)\mathcal H_0 +  \tilde f_1(t) \mathcal H_1,
\end{equation}
where $\tilde f_1(t)=f_1(t) - i f_2(t) $. The functions $f_0(t)$ and $f_1(t)$ (being monotonic decreasing and increasing, respectively) satisfy the following conditions: $f_1(0) \gg f_0(0)$, if $f_0(0)\neq 0$, and $f_1(\tau)= 0$, $f_0(\tau) = 1$.
In addition, we assume that the function $f_2(t)$ is monotonic and $f_2(\tau)= 0$.

The evolution of the total system is determined by the Schr\"odinger equation and its adjoint equation \cite{GW}:
\begin{align}\label{NSch1}
i\frac{\partial }{\partial t}|\psi(t)\rangle &= {\mathcal {\tilde H}_\tau(t)}|\psi(t)\rangle, \\
-i\frac{\partial }{\partial t}\langle\tilde\psi(t)|& =
\langle\tilde\psi(t)|{\mathcal {\tilde H}_\tau(t)}\label{NS2}.
\end{align}
We impose the initial conditions as follows:
$\mathcal{\tilde H}_1|\psi_g\rangle= E_g|\psi_g\rangle$, $E_g$ being the energy of the ground state of the initial non-Hermitian Hamiltonian $\mathcal{\tilde H}_1 =   {\tilde f}_1(0) \mathcal H_1 $.

We denote by $|\psi_n(t)\rangle$ and $\langle\tilde\psi_n(t)|$ the right/left instantaneous eigenvectors of the total Hamiltonian:
\begin{align}
{\mathcal {\tilde H}_\tau(t)}|\psi_n (t)\rangle = E_n(t)|\psi_n(t)\rangle, \\ \langle\tilde\psi_n (t)| {\mathcal {\tilde H}_\tau(t)}=
    \langle\tilde\psi_n(t)|E_n(t).
\end{align}
We assume that both systems of left and right eigenvectors form a bi-orthonormal basis,
$\langle\tilde\psi_m(t)|\psi_{n}(t)\rangle = \delta_{mn}$ \cite{MF}.

For the non-Hermitian quantum optimization problem the criterion validity of adiabatic approximation can be written as \cite{GW,BHC,FAMN}
\begin{equation}\label{NQA3}
  \tau \gg \frac{\max |\langle \tilde \psi_e (t)|\mathcal {\dot {\tilde H}_\tau}(t)|\psi_g(t)\rangle|}{\min |E_e(t)- E_g(t)|^2}.
\end{equation}
This restriction is violated near the ground state degeneracy where the complex energy levels cross. The point of degeneracy is known as the exceptional point, and it is characterized by a coalescence of eigenvalues and their corresponding eigenvectors, as well. Therefore, studying the behavior of the system in the vicinity of the exceptional point requires special care \cite{KMS,SKM,MKS1}.

At the crossover point, $t_c$, we introduce the bi-orthonormal basis as follows: In the two-dimensional subspace spanned by the ground state and the first excited state of the total non-Hermitian Hamiltonian, ${\tilde H}_\tau$, we choose an orthonormal basis $\{|0\rangle, |1\rangle\}$ and complement it to the complete basis by adding the eigenvectors $|\psi_k(t_c)\rangle$ $(k= 2,\dots,N-1)$. Then the set of states $\{\langle 0|,\langle 1|, \langle \tilde\psi_k(t_c)|;|0\rangle, |1\rangle,|\psi_k(t_c)\rangle\}$, where $k$ runs from 2 to $N-1$, forms the bi-orthonormal basis of the $N$-dimensional Hilbert space.

Basically repeating the same procedure as for the Hermitian case, we obtain the effective two-dimensional non-Hermitian Hamiltonian as
\begin{equation}\label{EH2}
  \mathcal {\tilde H}_{ef}(t)= \tilde\lambda (t){1\hspace{-.125cm}1} +f_0(t)\boldsymbol R_0\cdot \boldsymbol \sigma + \tilde f_1(t)\boldsymbol R_1\cdot \boldsymbol \sigma,
\end{equation}
where $\tilde\lambda = f_0(t)\lambda_{0} + {\tilde f}_1(t)\lambda_{1} $, and $\boldsymbol R_0$ and $\boldsymbol R_1$ are defined in the same functional form as in Eq. (\ref{EH1}), but in the new basis referred to the crossover point of the total non-Hermitian Hamiltonian, $ \mathcal {\tilde H}_{\tau}(t)$.

The Hamiltonian $\mathcal {\tilde H}_{ef}$ has a complex energy spectrum given by $ E_{\pm} = \lambda \pm \tilde R$, where
\begin{equation}\label{R1c}
\tilde R =  \sqrt{{\tilde g}^2 - 2{\tilde g} J\cos\alpha +J^2},
\end{equation}
and we set $J=f_0(t)R_0$,  $\tilde g= g(t) - i \delta(t)={\tilde f}_1(t)R_1$. From (\ref{R1c}) we obtain the energy gap as
\begin{equation}\label{R2}
|\Delta  E | =  2|\sqrt{g^2 - 2 g J\cos\alpha +J^2- \delta^2 -2i\delta(g -  J \cos\alpha)}|.
\end{equation}
In Eqs. (\ref{R1c}), (\ref{R2}) and below we do not indicate in some expressions the explicit dependences on t.

As can be seen, $|\Delta  E|$ vanishes at the exceptional point, defined in the parameter space $(\delta, g)$ by
$g^2 + \delta^2 = J^2 , \quad
g=J\cos\alpha $.
This yields $\cos\alpha = \sqrt{1-(\delta/J)^2}$. From here it follows that the complex energy does not become degenerate during the evolution of the system, if  $\delta(t) > J(t)$ for any $0 \leq t \leq \tau$. Otherwise, an exceptional point appears for some time $t\leq \tau$.

From Eq. (\ref{R2}), taking into account that for many-qubits system $\sin \alpha \sim 2^{-n/2}$ and assuming that $\delta(t)/J(t) \gg 2^{-n/2} $ for any $0 \leq t \leq \tau$  , we obtain
\begin{align}\label{R5}
|\Delta E| \geq |\Delta E |_{\min}\approx 2\min(\sqrt{(g-J)^2 + \delta^2}),
\end{align}
where $n$ is the number of qubits. As can be seen, for any moment of time, $0 \leq t \leq \tau$, the minimum energy gap $|\Delta E |_{\min} \approx 2\min(\sqrt{(g-J)^2 + \delta^2}$. The complex energy gap is controlled by the parameter $\delta(t)$, and, thus, the non-Hermitian adiabatic optimization does not suffer from the typical exponentially small energy gap of the Hermitian adiabatic optimization.

A rough estimate of the time required for the non-Hermitian quantum optimization can be obtained applying the criterion of Eq. (\ref{NQA3}) to the effective Hamiltonian $\mathcal {\tilde H}_{ef}$. The estimation of the matrix element ${|\langle \tilde \psi_e (t) |\mathcal {\dot {\tilde H}_\tau}(t)|\psi_g(t) \rangle|}$ yields
\begin{align}\label{M1}
    {\max |\langle \tilde \psi_e (t)  |\mathcal {\dot {\tilde H}_\tau}(t)|\psi_g (t) \rangle|} \approx \frac{\max|J\dot {\tilde g} -\tilde g \dot J|\sin\alpha}{|\Delta E|_{\min}}.
\end{align}
From here, with help of Eq. (\ref{NQA3}) and taking into account  that for many qubits system $\sin\alpha \approx 2^{-n/2}$, we obtain
\begin{align}\label{R6}
\tau \gg \frac{2^{-n/2}\max|J\dot {\tilde g} -\tilde g \dot J|}{|\Delta E|_{\min}^3}.
\end{align}
\begin{figure}[tbh]
\begin{center}
\scalebox{0.45}{\includegraphics{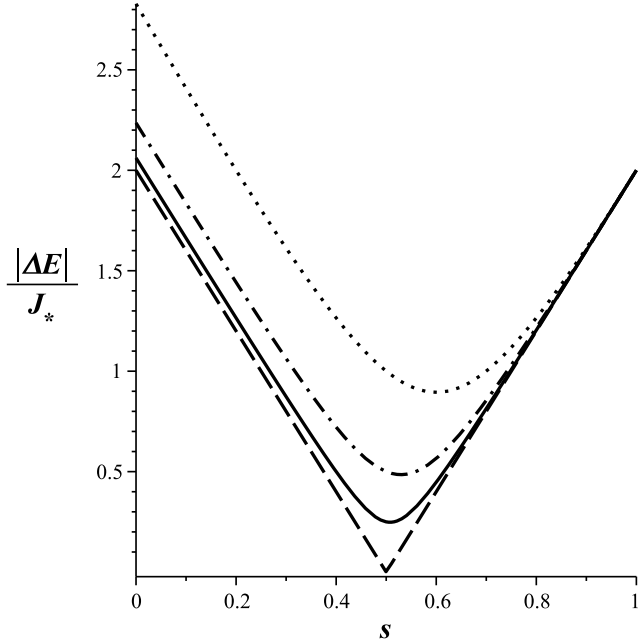}}
\end{center}
\caption{Non-Hermitian adiabatic quantum optimization ($\delta_0 \neq 0$): $|\Delta  E|/J_{\ast}$ as function of the dimensionless time $s=t/\tau$. The energy gap is renormalized and remains non-exponential for any time $0\leq s \leq 1$ (solid line, $\delta_0 =0.25J_{\ast}$; dash dotted line, $\delta_0 =0.5 J_{\ast}$; dotted line, $\delta_0 =J_{\ast}$). Hermitian adiabatic quantum optimization (dashed line, $\delta_0 =0$): The energy gap between the ground state and the first excited state becomes exponentially small at the moment of time $s=1/2$.  }
\label{NAQC1}
\end{figure}

As an illustrative example, we consider the non-Hermitian adiabatic quantum optimization algorithm realized by the following linear interpolation: $\tilde g(s) = g(s)- i\delta(s)=(J_{\ast}- i\delta_0)(1-s)$ and $J(s) = J_{\ast} s$, where $s=t/\tau$ denotes dimensionless time. Substituting $g(s)$ and $J(s)$ into Eq. (\ref{R5}), we find that the energy gap, $|\Delta E |$, is bounded from below by $|\Delta E |_{\min} = {2 J_{\ast} \delta_0}/{\sqrt{\delta_0^2 + 4J_{\ast}^2}}$. At the critical point the complex energy has non vanishing gap controlled by the parameter $\delta_0$. (See Fig. \ref{NAQC1}.)
Applying the criterion of validity of the adiabatic approximation (\ref{NQA3}), we find that the evolution time $\tau$ must satisfy the following condition:
$ \tau \gg\tau_0 = 2^{-n/2} J_{\ast}{(\delta_0^2 + J_{\ast}^2)^{1/2}}/{|\Delta E |^3_{\min}}$.
From here, in the limit $\delta_0 \ll J_\ast$, we obtain $\tau \gg 2^{-n/2}J^2_\ast/\delta^3_0$.

This inequality means that the time of passing the energy gap point becomes non-exponentially small when the number of qubits, $n$, is big enough. There are two main reasons for this: (i)  due to the non-Hermitian part of the total Hamiltonian, the energy gap is renormalized and becomes non-exponential, and (ii) at the same time, the matrix element in (\ref{M1}) of transition between the ground state and the first excited state still remains exponentially small. We should also mention that the non-Hermitian part of the Hamiltonian leads to the final life-time of qubits. So, the following inequality must be satisfied: $2^{-n/2}J^2_\ast/\delta^3_0\ll\tau \ll 1/\delta_{qubit}$, where $\delta_{qubit}$ is the charcteristic width of the intrinsic energy level(s) of the individual qubit. We should note that the relation between $\delta_{qubit}$  and $\delta_{0}$  will require a concretization of the quantum computer architecture in (\ref{NQA4}).

\section{Concluding remarks}

In conclusion, we have demonstrated that our  adiabatic quantum optimization algorithm, based  on the use of non-Hermitian Hamiltonians,  can significantly reduce the time needed for optimization of complex combinatorial problems. The main ideas of our approach are the following. One uses as an individual qubit a two level system which has simultaneously the discrete (intrinsic) states and a continuous part of the spectrum. So, the Feshbach projection on the intrinsic states results in the finite widths, $\delta_{qubit}$,  for of the intrinsic states, which characterize the strength of the interaction of a qubit with a continuum. This interaction should not be too strong, as it characterizes the life-time of a qubit, $\sim 1/\delta_{qubit}$. On the other hand, the parameter, $\delta_{0}$, which characterizes the non-Hermitian part of the total Hamiltonian is responsible for the renormalization of the energy gap between the ground state and the first excited state. Then, the energy gap becomes non-exponential, but the matrix element which is responsible for a transition between the ground state and the first excited state still remains exponentially small. All these factors, taken together, result in the improving of the adiabatic condition, if the number of qubits is big enough. The possible candidates for implementation of this approach could be, for example, the superconducting phase qubits \cite{SAM,SMC,YHC,KNAB,MJM,BN1} which are effectively used in recent experiments on quantum computation. In order to establish the explicit relation between the parameters $\delta_{qubit}$ and $\delta_{0}$ one needs to build a concrete architecture of the quantum computer model, which in this paper is represented in a general form by Eq. (\ref{NQA4}).

\section*{Acknowledgments}

This work was carried out under the auspices of the National Nuclear
Security Administration of the U.S. Department of  Energy at Los
Alamos National Laboratory under Contract No. DE-AC52-06NA25396, and research grant SEP-PROMEP 103.5/04/1911.


\end{document}